\documentclass[%
superscriptaddress,
showpacs,
preprint,
showpacs,preprintnumbers,
amsmath,amssymb,
aps,
endfloats,
]{revtex4-1}

\usepackage[dvipdfmx]{graphicx}
\usepackage{dcolumn}
\usepackage{bm}
\usepackage{multirow}
\usepackage[mathlines]{lineno}

\usepackage{color}

\begin{document}

\preprint{APS/123-QED}

\title{Quantitative evaluation of Dirac physics in PbTe}

\author{Kazuto~Akiba}
\email{akb@okayama-u.ac.jp}
\altaffiliation[Current affiliation: ]{Department of Physics, Okayama University, Okayama 700-8530, Japan}
\affiliation{
The Institute for Solid State Physics, The University of Tokyo, Kashiwa, Chiba 277-8581, Japan
}

\author{Atsushi~Miyake}
\affiliation{
The Institute for Solid State Physics, The University of Tokyo, Kashiwa, Chiba 277-8581, Japan
}

\author{Hideaki~Sakai}
\affiliation{
Department of Physics, Osaka University, Toyonaka, Osaka 560-0043, Japan
}
\affiliation{JST-PRESTO, Kawaguchi, Saitama 332-0012, Japan}

\author{Keisuke~Katayama}
\affiliation{
Department of Physics, Osaka University, Toyonaka, Osaka 560-0043, Japan
}

\author{Hiroshi~Murakawa}
\affiliation{
Department of Physics, Osaka University, Toyonaka, Osaka 560-0043, Japan
}

\author{Noriaki~Hanasaki}
\affiliation{
Department of Physics, Osaka University, Toyonaka, Osaka 560-0043, Japan
}

\author{Sadao~Takaoka}
\affiliation{
Department of Physics, Osaka University, Toyonaka, Osaka 560-0043, Japan
}

\author{Yoshiki~Nakanishi}
\affiliation{
Faculty of Science and Engineering, Graduate School of Engineering, Iwate University, Morioka, Iwate 020-8551, Japan
}

\author{Masahito~Yoshizawa}
\affiliation{
Faculty of Science and Engineering, Graduate School of Engineering, Iwate University, Morioka, Iwate 020-8551, Japan
}

\author{Masashi~Tokunaga}
\affiliation{
The Institute for Solid State Physics, The University of Tokyo, Kashiwa, Chiba 277-8581, Japan
}

\date{\today}

\begin{abstract}
The magnetic field dependence of electronic transport, magnetic, and elastic properties
in single crystals of PbTe were investigated in high magnetic fields up to 55 T.
The magnetoresistance, magnetization, ultrasonic attenuation,
and sound velocity showed clear quantum oscillations
with pronounced Zeeman-splitting,
which causes a large second harmonic in the frequency spectra.
The ratio of the Zeeman to the cyclotron energy, which is regarded
as an index of ``Diracness'' [H. Hayasaka and Y. Fuseya, J. Phys.: Condens. Matter \textbf{28,} 31LT01 (2016).],
was determined to be 0.52 and 0.57 in samples with different carrier densities.
We also pointed out that the effect of Zeeman-splitting seriously affects the Landau-level fan diagram analysis,
which is widely used to extract the nontrivial Berry's phase from the quantum oscillations.
\end{abstract}

\maketitle


\section{introduction}
\label{sec_intro}
Materials with nontrivial band topology are called topological materials, 
which do not belong to the categories based on conventional band theory.
The exploration of novel physical properties in these materials
is now one of the main streams in condensed matter physics.
One of the outstanding features of topological materials is the presence of
Dirac fermions, which obey the relativistic Dirac equation and have extremely high mobility.
The first prediction and experimental discovery of topological insulators \cite{Fu_topo, Hsieh_topo},
which possess Dirac fermions at their surfaces, initiated the research on the topological nature of solids.
More recently, three-dimensional (3D) topological semimetals such as Dirac semimetals \cite{Liu_Cd3As2, Liu_Na3Bi}
and Weyl semimetals \cite{Xu_TaAs, Wu, Deng}
have been actively investigated owing to their exotic character,
in which Dirac fermions contribute
to their bulk physical properties.

The identification of topological materials relies on experimental or computational
verification of the linear energy dispersion relation,
which is a consequence of the Dirac equation.
In fact, in 3D topological semimetals such as Cd$_3$As$_2$ and Na$_3$Bi,
the linearity of the dispersion relation has been confirmed
by both ARPES experiments and band calculations \cite{Neupane, Liu_Na3Bi, Wang_Na3Bi}.
Other materials have been proposed as candidates for topological materials on the basis of the above criteria.

The point of interest in these materials is the novel transport properties that emerge
as a result of the relativistic equation of motion in these materials.
One of the intriguing properties expected in topological materials is the nontrivial Berry's phase
accompanied by the singularity of the energy band structure.
Mikitik \textit{et al.} theoretically suggested that
a closed orbit in the momentum space that surrounds a Dirac point in a two-dimensional (2D) surface state
or band-contact line in a 3D bulk
has a nontrivial Berry's phase $\Phi_B=\pi$ \cite{Mikitik}.
They also pointed out that whether a material has a nontrivial Berry's phase
can be identified by focusing on the phase of the quantum oscillations.
A widely used method to estimate $\Phi_B=\pi$ is called Landau-level fan diagram analysis \cite{Ando_JPSJ, Murakawa}.
In the case of the Shubnikov--de Haas (SdH) oscillation,
the conductivity $\sigma_{xx}$ is plotted against the inverse of the magnetic field ($1/B$).
Then, a \textit{peak} position ($1/B_n$) is assigned to an \textit{integer} Landau index $n$.
Through the above assignment, a Landau-level fan diagram ($1/B_n$ against $n$) is constructed.
Here, the oscillation component of the SdH oscillation $\Delta \sigma_{xx}$ is assumed to be
\begin{equation}
\Delta \sigma_{xx}\propto \cos \left[2\pi \left(\dfrac{F}{B}+\gamma\pm\delta \right) \right],
\label{eq_fan_plot}
\end{equation}
where $F$ is the frequency of oscillation.
$\gamma=1/2-\Phi_B/(2\pi)$ is called the Onsager phase factor with Berry's phase $\Phi_B$,
which originates from the Bohr--Sommerfeld quantization rule for a closed orbit in the momentum space
surrounding the area $S_n$:
\begin{equation}
S_n=\dfrac{2\pi eB}{\hbar}(n+\gamma),
\end{equation}
where $e$ and $\hbar=h/2\pi$ are the elemental charge and reduced Planck's constant, respectively.
$\delta$ takes values of $\pm1/8$ and $0$ for 3D and 2D systems, respectively.
Eq. (\ref{eq_fan_plot}) shows that
the system has a nontrivial Berry's phase $\Phi_B=\pi$
when the horizontal intercept is $\pm1/8$ (3D bulk) or 0 (2D surface).
In contrast,
the trivial Berry's phase $\Phi_B=0$ is identified when the horizontal intercept takes $-1/2 \pm 1/8$ (3D bulk) or $-1/2$ (2D surface).
In many studies, the nontrivial Berry's phase detected through the phase of quantum oscillations
has been proposed as evidence of a topological material.
However, the precise determination of the phase factor of quantum oscillations is difficult because
most of the actual materials have multiple types of carriers. 
Owing to the complexity accompanying the analyses,
there exist several cases where the estimated Berry's phase varies depending on research groups
even in the same material \cite{Wang_Berry}.

Further, there are many reports that suggest observation of unconventional
transport properties owing to the presence of Dirac fermions.
One example is a large linear transverse magnetoresistance (a magnetic field is applied perpendicular to the electric current) in high magnetic fields.
The observation of a large linear magnetoresistance, for example, in Cd$_3$As$_2$ \cite{Liang}
and many other materials, has been reported.
In \cite{Liang}, the possible contribution of topologically protected backscattering
has been suggested as a mechanism of the phenomenon.
Another example of an intriguing phenomenon is negative longitudinal magnetoresistance
(a magnetic field is applied parallel to the electric current).
This behavior is believed to stem from the Adler--Bell--Jackiw chiral anomaly (or merely chiral anomaly)
in quantum field theory \cite{Nielsen}.
We can find many reports that claim observation of a chiral anomaly,
for example, in Na$_3$Bi \cite{Xiong}.
On the other hand, linear magnetoresistance and negative longitudinal magnetoresistance
can occur without the nontrivial band topology \cite{Parish, Kisslinger, Arnold_TaP},
and it is difficult to distinguish the nontrivial contribution from the experimental results.
Considering the abovementioned situations, the universal nature of the topological materials has not fully been clarified at the moment.

Recently, another approach to evaluate similarity to the Dirac system was proposed:
it focuses on the ratio of the Zeeman energy to the
cyclotron energy (ZC ratio), which well reflects the degree of relativistic effects in materials \cite{Hayasaka}.
An ideal 3D Dirac electron system can be represented by a two-band
Dirac Hamiltonian, in which the ZC ratio is strictly unity
regardless of the magnetic field direction.
In real materials, the contribution from the other bands disturbs the ideal Dirac state,
appearing as a deviation of the ZC ratio from unity.
The ZC ratio can be experimentally determined from the quantum oscillation phenomena under
magnetic fields.
Thus, the ZC ratio can serve as a quantitative index to evaluate the degree of a Dirac system,
so to speak, the ``Diracness'' of materials. 
In fact, the ZC ratio is confirmed to be almost unity in electrons in bismuth,
a typical 3D Dirac system \cite{Fuseya_Dirac}.
Nevertheless, there have been a limited number of studies that focus on the ``Diracness'' or ZC ratio.
Therefore, a clear methodology for determining the ZC ratio from experimental data and careful
confirmation of its validity are needed.
To achieve this, the material investigated should have as simple an electronic structure as possible
to eliminate ambiguity and complexity in the analysis.

PbTe, the target of this study,
is known as a degenerate narrow gap semiconductor with the NaCl-type cubic crystal structure,
a low carrier concentration of 10$^{17-18}$ cm$^{-3}$, and a high mobility of $\sim$$10^5$ cm$^2$V$^{-1}$s$^{-1}$ \cite{Allgaier}.
The narrowest band gap is located at the $L$ points in the first Brillouin zone.
By substituting Pb for Sn, the energy gap at the $L$ point becomes narrow,
and thus band touching is expected to be achieved at $x\sim0.35$ \cite{Dimmock}.
In composition with $x>0.35$, the band gap again opens due to band inversion, and the topological crystalline insulator phase
is realized where metallic surface energy bands protected by the crystal symmetry exist inside
the bulk band gap \cite{Xu_PbSnTe}.
A recent theoretical study predicted that the ZC ratio becomes unity at the topological phase transition
point \cite{Hayasaka}, indicating the possible realization of a 3D Dirac system.
A similar topological phase transition is predicted by applying hydrostatic pressure \cite{Barone}.
In PbTe, the physical properties are expected to be governed only by the carriers
in the vicinity of the $L$ point.
This simplicity is unlike other topological materials with multiple types of carriers.

In this article, we report the quantum oscillations observed in pristine PbTe at ambient conditions
probed by various measurements such as
resistivity, magnetization, and elastic properties under magnetic fields.
We observed clear quantum oscillations with pronounced Zeeman-splitting in all the measured physical quantities.
These results allowed us to determine the precise ZC ratio and Landau indices of quantum oscillations, 
which paved the way for an investigation tuning the ``Diracness'' with Sn substitution and
hydrostatic pressure.

\section{Experiments}
Single crystals of PbTe investigated in this study were prepared by the vapor transport
(referred to as the \#V sample) or Bridgman (referred to as the \#B sample) method.
As shown later, the carrier density of the \#V sample is approximately three times larger than that of the \#B sample.
Resistivities up to 14 T and magnetizations up to 7 T were measured with
the Physical Properties Measurement System (PPMS) and Magnetic Properties Measurement System (MPMS),
Quantum Design.
The magnetoresistance and Hall resistance of samples were simultaneously measured
by the standard five-probe method in samples with average dimensions of
2.0$\times$0.5$\times$0.2 mm$^3$
.
For magnetization measurements, we used samples with a mass of 139 mg for MPMS and a mass of 50.6 mg for pulsed magnetic fields.
Measurements up to 55 T were carried out using nondestructive pulse magnets
with a time duration of 36 ms at The Institute for Solid State Physics, The University of Tokyo.
Transport properties under pulsed magnetic fields were measured by a numerical lock-in technique
at typical AC frequencies of 100 kHz. 
Magnetization measurements in pulsed magnetic fields were carried out by the induction method using
two pick-up coils placed coaxially.
The ultrasonic attenuation coefficient and sound velocity were measured in pulsed magnetic fields
by a numerical implementation of the ORPHEUS method \cite{Sun, Fujita}.
A pair of LiNbO$_3$ transducers were attached to the cleavaged [001] planes of both sides of a
1.7-mm-long sample.
Ultrasound waves were injected by one transducer and detected by the other transducer.

\section{Results and Discussion}
\label{sec_rd}

\begin{figure}
\centering
\includegraphics[width=7.5cm]{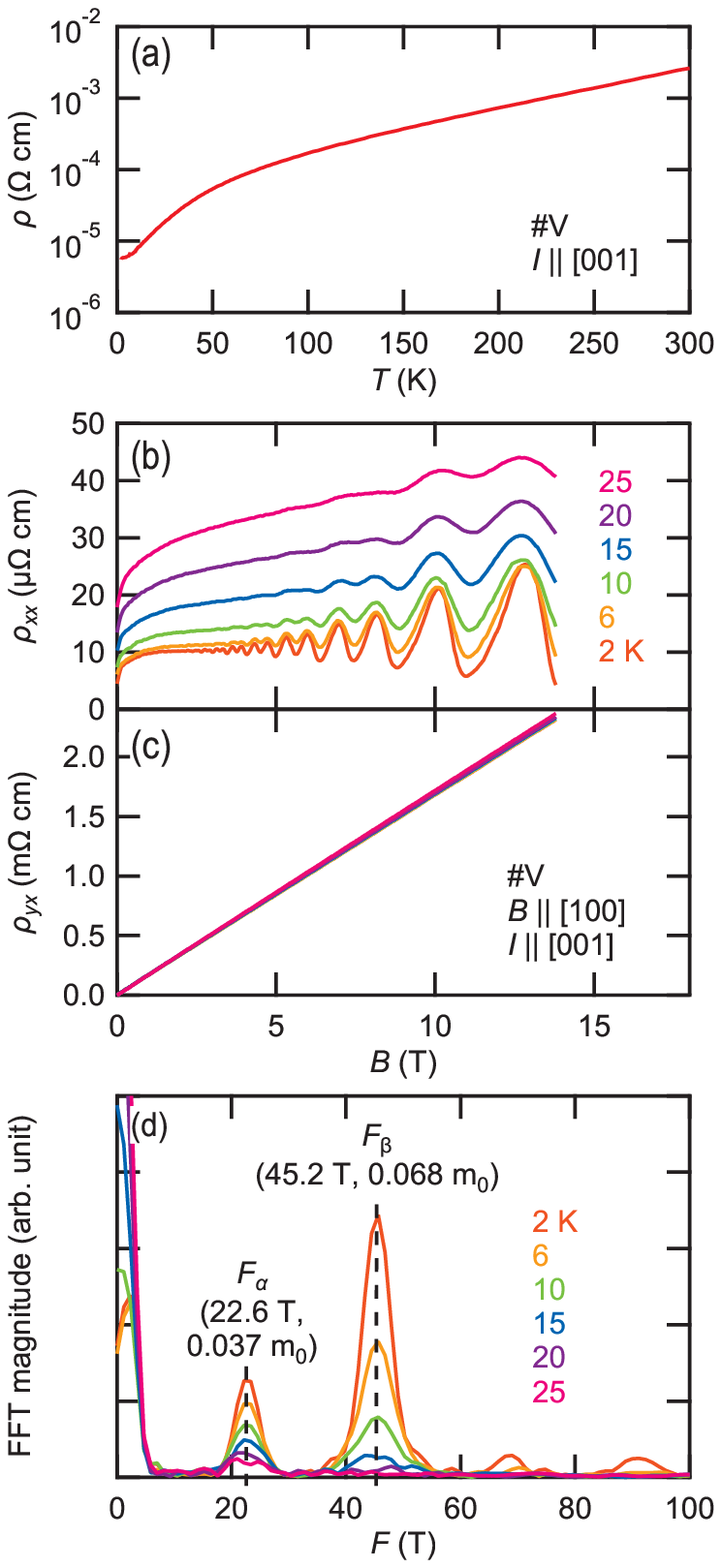}
\caption{
(a) Temperature dependence of the resistivity $\rho$ from 2 K to 300 K.
(b) Magnetoresistance $\rho_{xx}$ and
(c) Hall resistance $\rho_{yx}$ up to 14 T.
(d) FFT spectrum of SdH oscillation superimposed on $\rho_{xx}$.
\label{fig1}}
\end{figure}

First, we show the transport properties of the \#V sample.
Figure \ref{fig1}(a) shows the temperature dependence of the resistivity ($\rho$) with electric currents
applied along one of the principal axes of the cubic cell.
The resistivity showed metallic dependence in the entire temperature range from 300 to 2 K,
which indicates that the sample has a finite number of degenerated carriers at low temperature.
The RRR [$\rho$(300 K)/$\rho$(2 K)] is obtained as 453 from the data shown in Fig. \ref{fig1}(a).
Figure \ref{fig1}(b) shows the magnetoresistance ($\rho_{xx}$) up to 14 T at various temperatures.
In the following resistivity measurements, magnetic fields and currents were applied along the [100] and [001] directions, respectively, unless otherwise specified.
Although [100], [010], and [001] are symmetrically equivalent, we use these notations to
specify the relationship between the applied magnetic field and the electric current.
SdH oscillations are clearly observed at 2 K and are damped with increasing temperature.
Figure \ref{fig1}(c) shows the magnetic field dependence of the Hall resistivity ($\rho_{yx}$).
$\rho_{yx}$ is positive and increases linearly as the magnetic flux density ($B$) increases up to 14 T,
which suggests that a single type of hole carrier is responsible for the transport properties.
The slope of $\rho_{yx}$ is almost independent of temperature up to 25 K, which indicates that the hole density
($n_p$) is insensitive
to temperature.
It is notable that $\rho_{yx}$ is almost 100 times larger than $\rho_{xx}$.
Hence, SdH oscillations are inconspicuous in $\rho_{yx}$ owing to the huge linear background.
The hole density ($n_p=3.7\times 10^{18}$ cm$^{-3}$) and mobility ($\mu_p=3.8 \times 10^5$ cm$^2$V$^{-1}$s$^{-1}$) are estimated by the single carrier Drude model:
\begin{align}
\rho_{xx}&=\dfrac{1}{e\mu_p n_p},\\
\rho_{yx}&=\dfrac{B}{n_p e}.
\end{align}
Here, we evaluated $\mu_p$ using $\rho$ at 2 K.
The obtained $n_p$ and $\mu_p$ are consistent with the previous report \cite{Jensen}.

Here, we focus on the frequency of the SdH oscillations superimposed on $\rho_{xx}$.
Figure \ref{fig1}(d) shows the fast Fourier transform (FFT) spectra of the SdH oscillations.
We can recognize two obvious peaks (labeled $F_\alpha$ and $F_\beta$) showing systematic damping as the temperature increases.
Small cyclotron masses of 0.037 $m_0$ for $F_\alpha$ and 0.068 $m_0$ for $F_\beta$ were identified
from the temperature dependence of the SdH oscillation.
The existence of two types of Fermi pockets with different cross sections may result in this two-peak feature
with different cyclotron masses,
while it is inconsistent
with the single-carrier-like $\rho_{yx}$ shown in Fig. \ref{fig1}(c).
Because $F_{\beta} \simeq 2 F_{\alpha}$, 
we can alternatively interpret $F_\beta$ as the second harmonic of $F_\alpha$.
In this case, a reasonable explanation is needed for why
amplitudes of higher harmonics are larger than those of the fundamental wave.

\begin{figure}
\centering
\includegraphics[width=7.5cm]{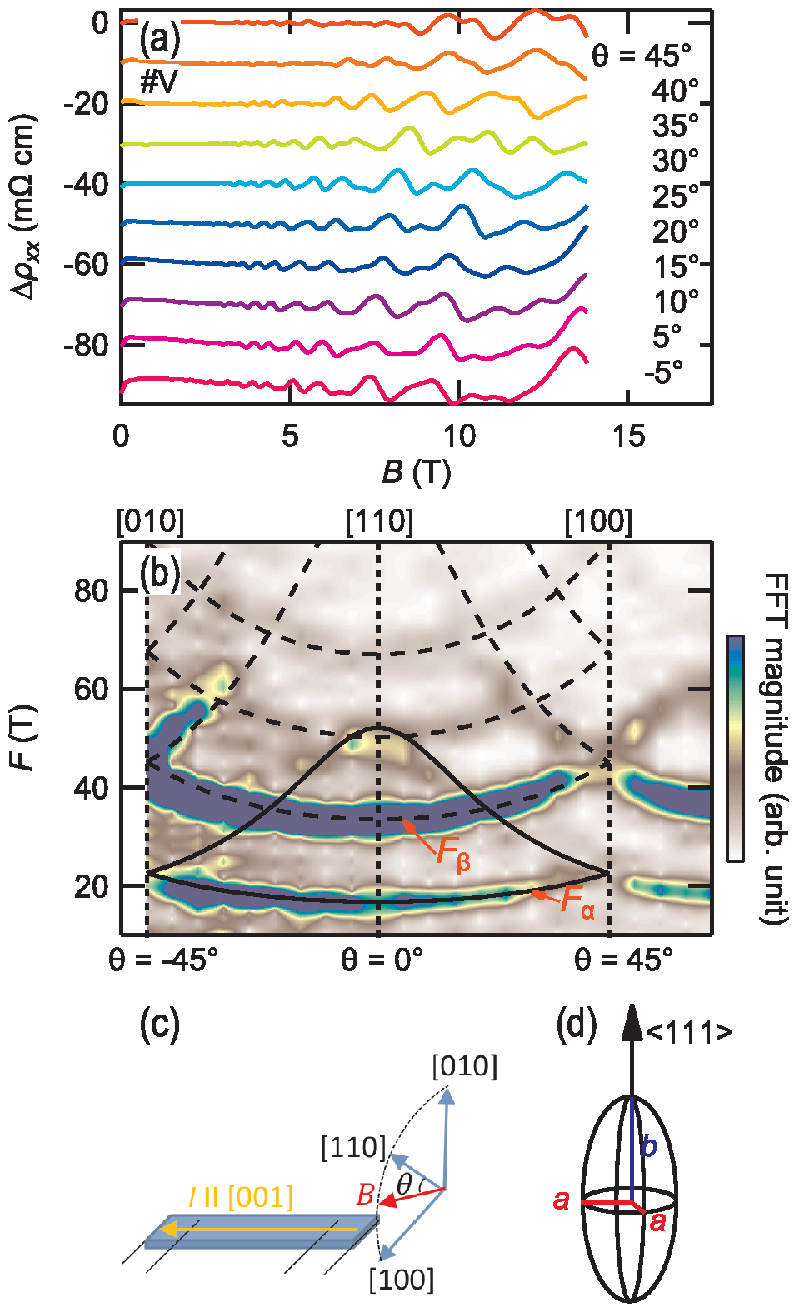}
\caption{
(a) Angular dependence of $\Delta \rho_{xx}$.
The magnetic field was swept along various directions in the plane perpendicular to the [001] direction.
$\theta$ represents the angle of $B$ from the [110] direction
as defined in (c).
(b) Contour plot of the FFT magnitude of the SdH oscillations in the frequency $F$ and field-angle plane.
Solid and dashed curves show the fundamental frequencies and higher harmonics, respectively,
calculated by assuming the ellipsoidal Fermi pocket as shown in (d) (see text in detail).
(c) Relationship between crystal axes, magnetic field, and current directions.
Magnetic fields are always applied perpendicular to the current.
(d) The ellipsoidal Fermi surface assumed in the analysis.
\label{fig2}}
\end{figure}

To obtain an insight into the origin of these frequency peaks, we investigated the angular dependence
of the SdH oscillations.
If the $F_\alpha$ and $F_\beta$ peaks originate from multiple Fermi pockets, these peaks vary independently.
On the other hand,
if $F_\beta$ is the second harmonic of $F_\alpha$, the relationship $F_\beta \simeq 2 F_\alpha$ should be
maintained regardless of the magnetic field direction.
Figure \ref{fig2}(a) shows the angular dependence of the oscillatory components ($\Delta \rho_{xx}$) superimposed on
$\rho_{xx}$.
Here, $\Delta \rho_{xx}$ was obtained by subtraction of the polynomial curve
from $\rho_{xx}$.
The magnetic field was tilted between the [100] and [110] directions in Fig. \ref{fig2}(c).
The SdH oscillation pattern varied upon the field direction in a complicated manner.
A detailed analysis and Landau indices for this oscillation pattern will be discussed later.
Here, we focus on the angular dependence of the oscillation frequencies.
Figure \ref{fig2}(b) shows the contour plot based on the FFT spectra calculated from $\Delta \rho_{xx}$
in various field directions.
In Fig. \ref{fig2}(b), the left axis and color bar represent the frequency
and amplitude of the SdH oscillations, respectively.
It is clearly shown that the relationship $F_\beta \simeq 2 F_\alpha$ is always satisfied, which supports the hypothesis
that $F_\alpha$ and $F_\beta$ originate from the same Fermi pocket at the $L$ point.
To quantitatively confirm this, we fitted the observed angular dependence of SdH frequencies 
with the single Fermi pocket model.
We assume that the Fermi pocket at the $L$ point is a simple ellipsoid aligned along the $\langle111\rangle$ direction,
which is characterized by the length of the short ($a$) and long ($b$) axes of the ellipsoid in the $k$-space
[Fig. \ref{fig2}(d)].
The solid lines overlayed on Fig. \ref{fig2}(b) show the angular dependence of the fundamental frequencies $F_{fund}$
calculated from the cross section of the ellipsoid perpendicular to the field direction.
In this calculation, we set the carrier density which is enclosed inside the ellipsoid to $1.1\times10^{18}$ cm$^{-3}$
and the anisotropy of ellipsoid to $K=(b/a)^2=13.7$.
This anisotropy factor reproduces the previous report \cite{Burke, Jensen}.
Because there are four ellipsoids in the first Brillouin zone, the total carrier density corresponds to
$4.4\times10^{18}$ cm$^{-3}$, which is in good agreement with $n_p=3.7\times 10^{18}$ cm$^{-3}$
determined from $\rho_{yx}$.
Moreover, we can reproduce the whole angular dependence of the SdH frequency up to 100 T by introducing
higher harmonics $2F_{fund}$, $3F_{fund}$, and $4F_{fund}$ [shown with broken lines in Fig. \ref{fig2}(b)],
which supports the validity of the above model.
The difference in the cyclotron masses between $F_\alpha$ (0.037 $m_0$)
and $F_\beta$ (0.068 $m_0$)
also supports this interpretation.
In ordinary cyclotron mass analysis,
the higher harmonics are sufficiently negligible compared to the fundamental harmonic.
If we apply the analysis to the peak of the 2nd harmonic,
$m^*$ is overestimated by a factor of two, which explains the difference in
cyclotron masses for $F_\alpha$ and $F_\beta$.
From these results, we conclude that Fermi pockets are located only at the $L$ points, and $F_\beta$ is 
the second harmonic of $F_\alpha$.

\begin{figure}
\centering
\includegraphics[width=7.5cm]{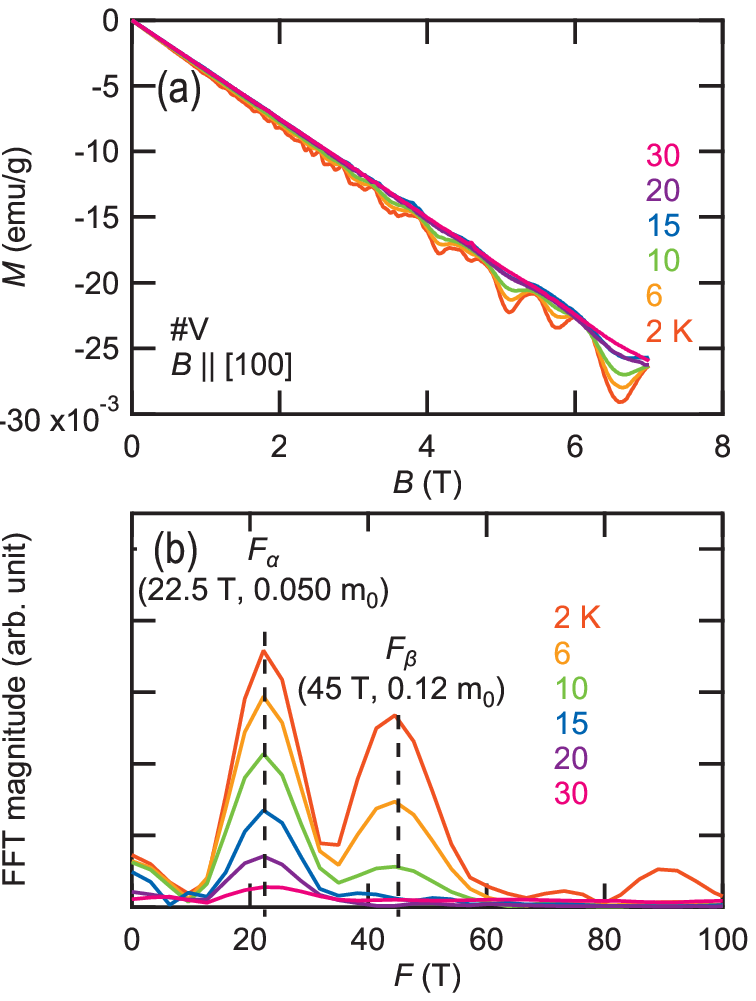}
\caption{
(a) Magnetic field dependence of the magnetization $M$ of \#V up to 7 T.
(b) FFT spectra of the dHvA oscillations of $M$.
\label{fig3}}
\end{figure}

Next, we focus on the magnetization ($M$), which is a thermodynamic quantity.
Figure \ref{fig3}(a) shows the magnetic field dependence of $M$ up to 7 T.
Here, $B$ was applied along the principal axis.
Clear de Haas--van Alphen (dHvA) oscillations overlap on the diamagnetic linear slope.
Figure \ref{fig3}(b) shows the FFT spectra of the dHvA oscillations.
The positions of the FFT peaks reproduce the result of transport measurements.
Because the maximum magnetic field is lower than that in the transport measurements, the magnitude of
the second harmonic is smaller than the fundamental harmonic.
The cyclotron masses differ
between $F_\alpha$ (0.050 $m_0$) and $F_\beta$ (0.12 $m_0$) approximately by a factor of 2,
which indicates that $F_\beta$ is the second harmonic of $F_\alpha$.

\begin{figure*}
\centering
\includegraphics[width=15cm]{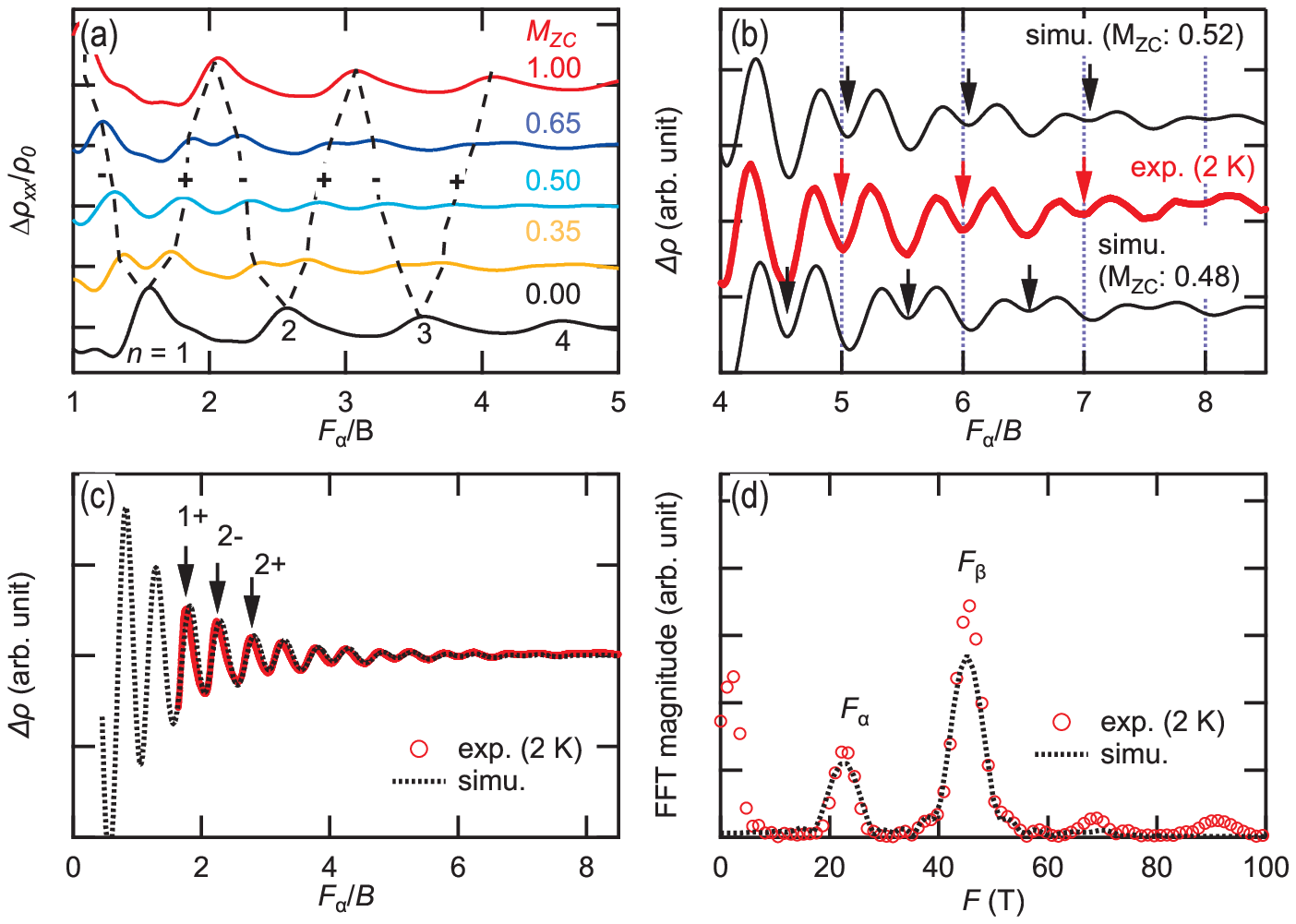}
\caption{
(a) Simulated oscillatory components in $\rho_{xx}$ as a function of $F_{\alpha}/B$
assuming $M_{ZC}=0.00$, 0.35, 0.50, 0.65, and 1.00 in Eq. (\ref{eq_LK}).
(b) Comparison of oscillatory components between the experiment at 2 K (red) and simulation (black)
in magnetic field region $F_\alpha/B>4$.
Simulation was performed for cases of $M_{ZC}=0.52$ (upper) and 0.48 (lower).
Positions of the shallower dip are indicated by arrows.
(c) Comparison of the whole SdH oscillation between the experiment at 2 K (red) and simulation (black).
(d) Comparison of the FFT spectra between the experimental data at 2 K (red symbols)
and simulated data (black broken line).
\label{fig4}}
\end{figure*}

Here, we consider the reason why the amplitude of $F_{\beta}$ becomes larger than that of $F_{\alpha}$,
by taking the effect of Zeeman-splitting into account.
According to the LK theory, the oscillatory component of $\rho_{xx}$ is represented
by the following form including the ZC ratio $M_{ZC}=g^*\mu_B B/(\hbar \omega_c)$ \cite{RA, Murakawa}:
\begin{equation}
\begin{split}
\dfrac{\Delta \rho_{xx}}{\rho_0} &\simeq \dfrac{5}{2} \sum_{p=1}^{\infty}
\sqrt{\dfrac{B}{2pF_{\alpha}}}\\ \times &R_T (T, p) R_D(T_D, p) R_S (M_{ZC}, p)\\
\times &\cos \left[ 2\pi p\left(\dfrac{F_{\alpha}}{B}+\dfrac{1}{2} \right)-\dfrac{\pi}{4} \right].
\label{eq_LK}
\end{split}
\end{equation}
Here, $\rho_0$ and $T_D$ are the resistivity at $B=0$
and the Dingle temperature, respectively.
The natural number $p$ represents the number of harmonics.
The phase factor $\gamma$ in Eq. (\ref{eq_fan_plot}) is $1/2$ because PbTe is assumed to be a trivial
semiconductor.
In addition, the Fermi surface of PbTe was found to be a simple ellipsoid,
and thus the double sign in the third
term inside the cosine function in Eq. (\ref{eq_fan_plot}) was set to be negative.
Each cosine term has three damping factors, namely,
the temperature factor $R_T(T, p)$:
\begin{equation}
R_T(T, p)=\dfrac{2\pi ^2 pk_B T/(\hbar \omega_c)}{\sinh [2\pi ^2 pk_B T/(\hbar \omega_c)]},
\end{equation}
the Dingle factor $R_D(T_D, p)$:
\begin{equation}
R_D(T_D, p)=\mathrm{exp}\left(-\dfrac{2\pi ^2 pk_B T_D}{\hbar \omega_c}\right),
\end{equation}
and the spin factor $R_S(M_{ZC},p)$:
\begin{equation}
R_S(M_{ZC},p)=\cos(pM_{ZC}\pi).
\end{equation}
We consider a case where $g^*\mu_B B$ is equal to half of $\hbar \omega_c$, that is, $M_{ZC}=0.5$.
In this case, 
$R_S(0.5, p)=0$ for odd $p$, while $R_S(0.5, p)=\pm 1$ for even $p$.
Although $R_T$ and $R_D$ exponentially decrease with increasing $p$,
the second harmonics can dominate the fundamental harmonic
when $M_{ZC}$ is close to 0.5. 
Therefore, we can expect a relatively large second harmonic for $M_{ZC}$ close to $0.5,1.5,2.5,\cdots$.
Figure \ref{fig4}(a) shows the simulated modification of the SdH oscillation by incrementing $M_{ZC}$ from 0 to 1.
Each curve is simulated substituting appropriate values of $m^*$,
$F_{\alpha}$, and $T_D$ in Eq. (\ref{eq_LK}).
The peaks of resistivity initially split into two as indicated by $+$ and $-$ in Fig. \ref{fig4}(a),
and move in opposite directions as $M_{ZC}$
increases toward 1.
Then, at $M_{ZC}=1$, two peaks having different Landau indices merge into the same peak.
Here, the phase of oscillation differs by $\pi$ compared to that in $M_{ZC}=0$.
As expected above,  the period of the oscillation at $M_{ZC}=0.5$ becomes just half of that at $M_{ZC}=0$
or 1.
We then compare our experimental results with the simulated SdH oscillation around $M_{ZC}=0.5$,
and find a suitable $M_{ZC}$.
Figure \ref{fig4}(b) shows the comparison between the experimental result at 2 K (red)
and the simulated curves (black) based on Eq. (\ref{eq_LK}).
These curves are vertically shifted for clarity.
A simulation was performed for $M_{ZC}=0.52$ and 0.48 assuming parameters of
$m^*=0.05$ $m_0$, $T_D=10$ K, and $F_{\alpha}=22.6$ T.
Relatively small contributions from $p\geq3$ are ignored.
Although the waveforms of $M_{ZC}=0.52$ and $M_{ZC}=0.48$
look quite similar,
we can see that the simulated curve with $M_{ZC}=0.52$ better reproduces the experimental result
than that with $M_{ZC}=0.48$
[focus on the positions of the shallower dips indicated by arrows in Fig. \ref{fig4}(b)].
Figure \ref{fig4}(c) shows the comparison of the whole SdH oscillation between the experimental result at 2 K (red)
and the simulated curves (black) with $M_{ZC}=0.52$.
Both the peak/dip structures and the phase of the oscillation
agree well with the experimental result.
The Landau indices of the last three peaks are shown assuming $M_{ZC}=0.52$.
Figure \ref{fig4}(d) shows the FFT spectrum of the experimental data at 2 K (red circles) and
the simulated curve (broken line).
We can reasonably reproduce the magnitude ratio between $F_\alpha$ and $F_\beta$
and the anomalously large second harmonics.

\begin{figure}
\centering
\includegraphics[width=7.5cm]{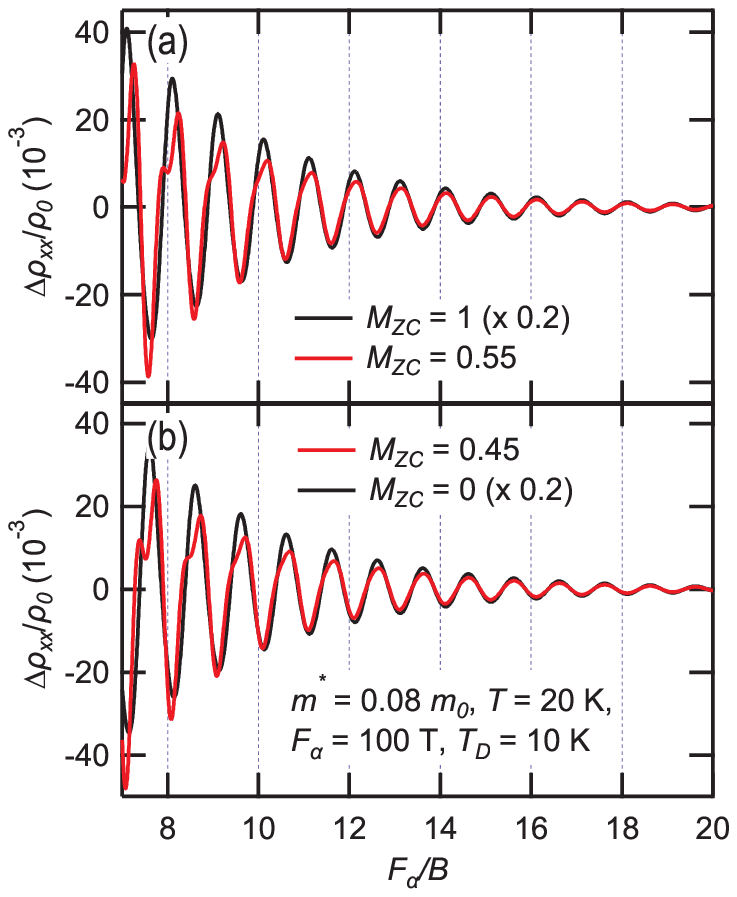}
\caption{
Comparison of the simulated oscillation structures between (a) $M_{ZC}=1$ and $M_{ZC}=0.55$ cases,
and (b) $M_{ZC}=0.45$ and $M_{ZC}=0$ cases.
Simulations were performed on the basis of Eq. (\ref{eq_LK}), assuming $m^*=0.08 m_0$, $T=20$ K,
$F_\alpha=100$ T, and $T_D=10$ K.
\label{fig5}}
\end{figure}

Here, we note that whether $M_{ZC}$ is greater or less than 0.5
can have a crucial influence on the construction of the Landau-level fan diagram
under a specific condition.
As mentioned in Sec. \ref{sec_intro}, observation of the SdH oscillation and the construction of the Landau-level fan diagram
are frequently used methods
for the identification of the Berry's phase.
However, it is important to take the effect of the Zeeman-splittings into account.
In the general case, the spin factor $R_S=\cos(pM_{ZC}\pi)$ can affect the determination
of the phase.
We assume a trivial system ($\gamma=1/2$, $\Phi_B=0$) that has a light carrier mass of $m^*=0.08 m_0$, 
fundamental frequency of $F_\alpha=100$ T,
and moderate Dingle temperature of $T_D=10$ K.
We performed the numerical simulation for the above case on the basis of Eq. (\ref{eq_LK}).
The results are shown in Figs. \ref{fig5}(a) and (b).
Here, we assumed a relatively high temperature of $T=20$ K,
and showed the cases of $M_{ZC}=1$, 0.55, 0.45, and 0.
As is clear in Figs. \ref{fig5}(a) and (b), the Zeeman-splittings are discernible only for $F_\alpha/B<10$.
Thus, in both Figs. \ref{fig5}(a) and (b),
the peak/dip positions look identical at a glance in magnetic fields with $F_\alpha/B>10$.
That is, when the magnetic field is not strong for the Zeeman-splitting effect to be discernible,
the fan diagram analysis cannot distinguish $M_{ZC}=1$ and 0.55,
or $M_{ZC}=0.45$ and 0 in spite of the significant difference in the ZC ratio.
Further, we focus on the cases of $M_{ZC}=0.45$ and 0.55 in Fig. \ref{fig5}.
Although the difference in the ZC ratio is only 0.1 between these cases,
the phase of the oscillations differs by $\pi$.
Consequently, the fan diagram analysis may lead the conclusion that the nontrivial Berry's phase has been realized
even in trivial materials with $M_{ZC}=0.55$ when only the oscillation in the region $F_\alpha/B>10$ is available.
Therefore, we have to be careful in the fan diagram analysis
when the Zeeman-splitting is smeared out by the broadening of the Landau subbands.
High-field experiments for $F_\alpha/B<10$ are necessary to distinguish whether the Zeeman-split exists or not
in the above situation.

\begin{figure}
\centering
\includegraphics[width=7.5cm]{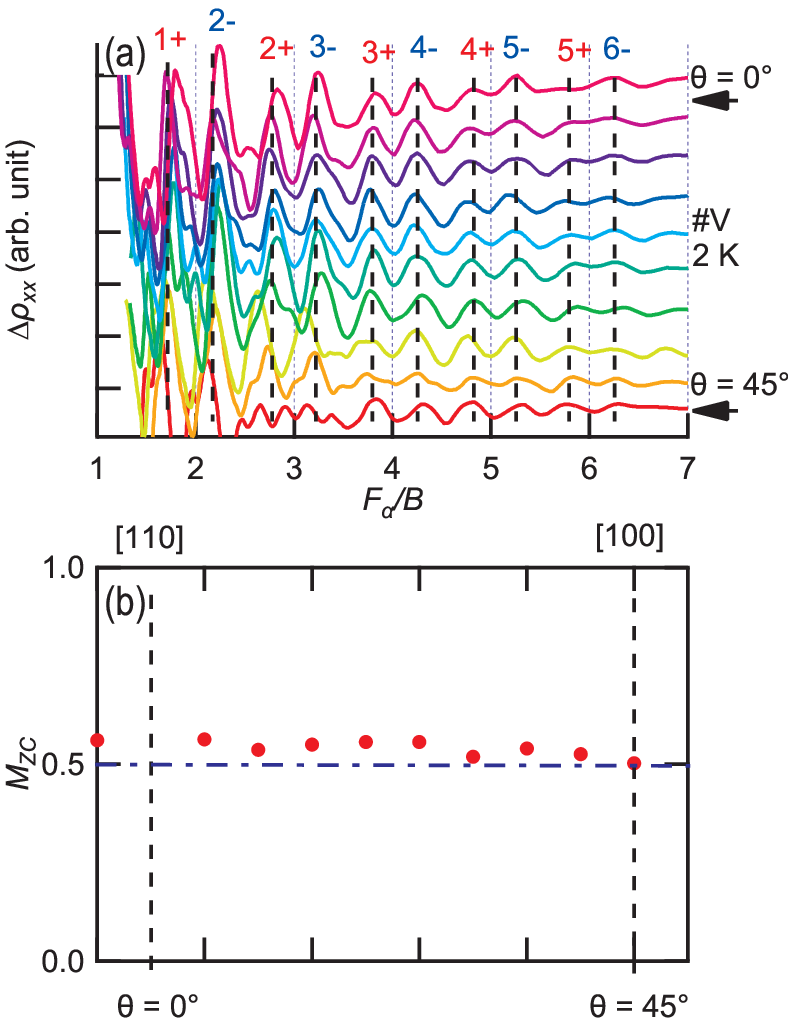}
\caption{
(a) $1/B$ dependence of $\Delta \rho_{xx}$
for $B$ at various angles between [100] and [110].
The horizontal axis was normalized by $F_\alpha$,
which was determined by the FFT of the SdH oscillation at each angle.
(b) Angular dependence of the ZC ratio from $\theta=0^\circ$ to $45^\circ$.
\label{fig6}}
\end{figure}

Next, we focus on the detailed structure of the angle-resolved SdH oscillations.
Figure \ref{fig6}(a) shows $\Delta \rho_{xx}$ with the magnetic field rotated from
[100] to [110] in the (001)-plane as shown in Fig. \ref{fig2}(c).
The horizontal axis is normalized by each $F_\alpha$ obtained by the FFT.
Starting from the Landau indices for $B\parallel [100]$ shown in Fig. \ref{fig4}(c),
we can identify the Landau indices at various field directions.
We note that significant anisotropy in $M_{ZC}$ has been observed in a previous report \cite{Burke}.
If $M_{ZC}$ is anisotropic,
the peak positions should vary with the field direction in Fig. \ref{fig6}(a).
However, our result shows that the peaks are almost independent of the field direction,
which indicates an isotropic $M_{ZC}$.
Here, we obtained the angular dependence of $M_{ZC}$ on the basis of the following equation \cite{Burke}:
\begin{equation}
\dfrac{(1/B)_{n+}-(1/B)_{n-}}{(1/B)_{n+}-(1/B)_{(n-1)+}}=M_{ZC}.
\label{eq_Burke}
\end{equation}
In Eq. (\ref{eq_Burke}), $(1/B)_{n+}$, \textit{etc.}, is the value of the inverse field at which
the energy level $n+$ crosses the Fermi level.
$M_{ZC}$ shown in Fig. \ref{fig6}(b) is the averaged value over the $n=5,4,3$ cases in
Eq. (\ref{eq_Burke}).
$M_{ZC}$ is almost independent of the field direction and exhibits an isotropic nature
in contrast to the previous report.
An isotropic $M_{ZC}$ is consistent with the recent theoretical prediction
by Hayasaka \textit{et al} \cite{Hayasaka}.

\begin{figure}
\centering
\includegraphics[width=7.5cm]{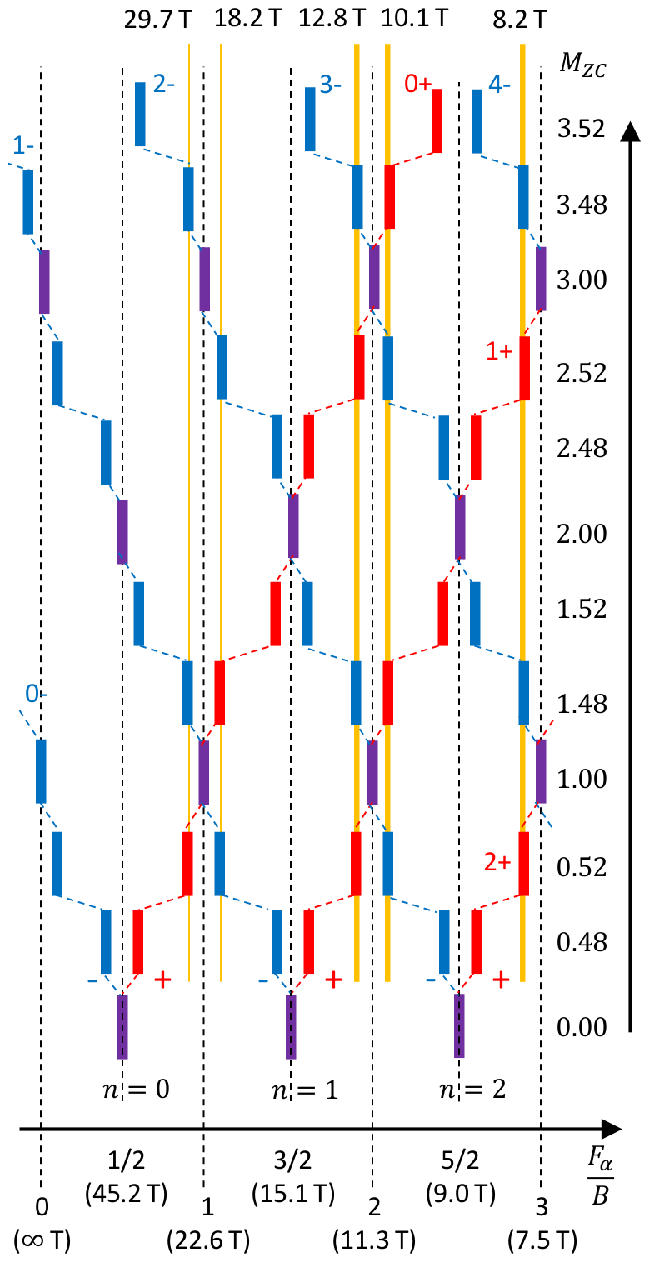}
\caption{
Schematic Landau-level structures at several ZC ratios.
The horizontal and longitudinal axis represent the values of $F_\alpha/B=n+1/2\pm M_{ZC}$ and $M_{ZC}$, respectively.
\label{fig7}}
\end{figure}

Up to here, we found that the oscillation structure was reproduced by assuming a fixed value of $M_{ZC}=0.52$,
independent of the field direction.
However, $M_{ZC}=0.52$ and the Landau indices shown in Figs. \ref{fig4}(c) and
\ref{fig6}(a) are not uniquely determined, as explained below.
Figure \ref{fig7} shows the schematic Landau-level structures at several $M_{ZC}$ up to 3.52.
Because the oscillation frequency was found to be $F_\alpha=22.6$ T,
all the level-crossing fields can be calculated by the relationship with $n=0,1,2\cdots$ \cite{LK_form}:
\begin{equation}
\dfrac{F_\alpha}{B}=n+\dfrac{1}{2} \pm \dfrac{1}{2}M_{ZC},
\label{eq_invpos}
\end{equation}
ignoring the dimensional factor $\delta$ in Eq. (\ref{eq_fan_plot}).
The observed peak positions in $\rho_{xx}$ up to 14 T are shown by the vertical bold orange lines.
The cases with $M_{ZC}>3.48$ apparently cannot explain the observed peak positions.
Among the cases shown in Fig. \ref{fig7},
we can see that 
$M_{ZC}=0.52, 1.48, 2.52,$ and 3.48 can reproduce the observed peak positions up to 14 T.
We cannot eliminate this ambiguity by the data up to 14 T.
To further restrict the possible cases,
experiments with higher magnetic fields are crucial.
If two oscillations are observed at approximately 18 and 30 T
(indicated by vertical thin orange lines in Fig. \ref{fig7}),
$M_{ZC}$ is 0.52 or 1.48.
If only one oscillation is observed,
$M_{ZC}$ is 2.52 or 3.48.
To clarify this point, we performed quantum oscillation measurements in pulsed magnetic fields
up to 55 T.

\begin{figure}
\centering
\includegraphics[width=7.5cm]{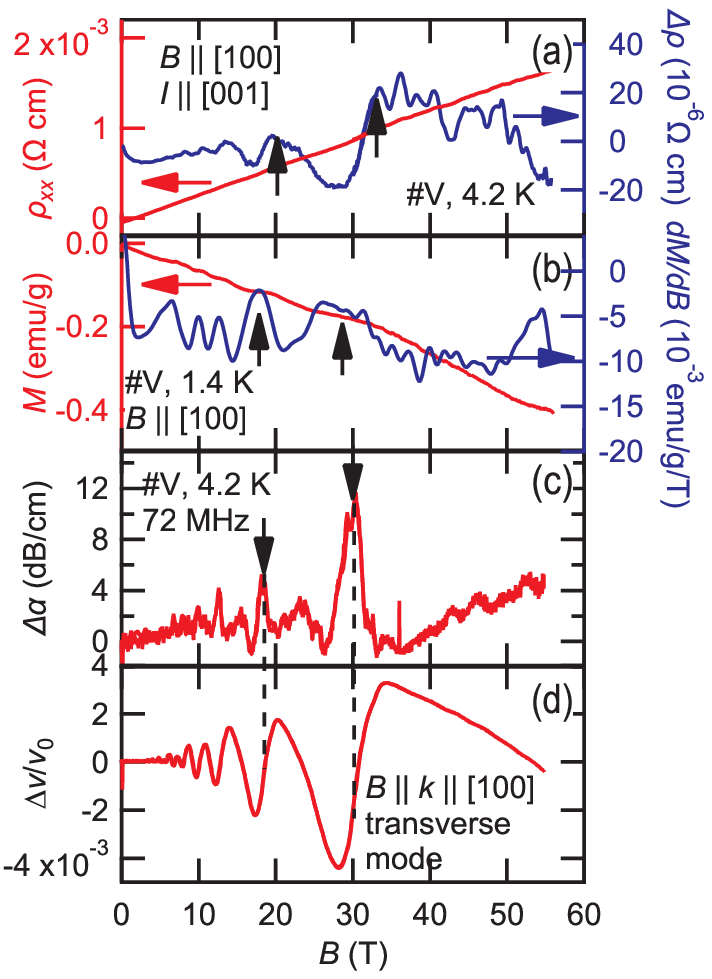}
\caption{
Magnetic field dependence of (a) magnetoresistance, (b) magnetization, (c) ultrasonic attenuation
coefficient, and (d) sound velocity up to 55 T.
In (a), blue trace is oscillatory components ($\Delta \rho$) obtained by subtracting the linear background from
the red trace.
In (b), blue trace is the first derivative of the red trace with respect to the magnetic field ($dM/dB$).
In (a) to (c), newly identified structures in high-field measurements are indicated by solid arrows.
\label{fig8}}
\end{figure}

Figures \ref{fig8}(a) and (b) show the magnetic field dependence of the resistivity and magnetization, respectively.
In Fig. \ref{fig8}(a), the red line indicate the raw field dependence of $\rho_{xx}$,
and the blue one $\Delta \rho$ obtained by subtraction of the linear background from $\rho_{xx}$.
In Fig. \ref{fig8}(b),
the red line indicate the raw field dependence of $M$,
and the blue one $dM/dB$ obtained by taking the first derivative of $M$ with respect to the magnetic field.
We observed two local maxima ($\sim$20 and 30 T) in both $\Delta \rho$ and $dM/dB$,
and no oscillation was found in higher fields up to 55 T.
In addition, the diamagnetism shown in Fig. \ref{fig3}(a)
was found to increase up to 55 T. 
We note that the field dependence of the magnetoresistance in $\rho_{xx}$ is different from that in
Fig. \ref{fig1}(a) measured in a different sample piece.
In spite of the difference in the underlying magnetoresistance,
the peak/dip positions of the SdH oscillations are reproduced well between the samples with
different magnetoresistance, and thus the quantum oscillation itself is almost independent of the sample piece.
Furthermore, we measured the elastic properties through ultrasound measurements in pulsed magnetic fields.
Figures \ref{fig8}(c) and (d) show the change in the
ultrasonic attenuation coefficient ($\Delta \alpha$) and sound velocity ($\Delta v/v_0$) from zero field, respectively, as a function of magnetic field up to 55 T.
$v_0$ represents the sound velocity at $B=0$.
In both traces, clear acoustic quantum oscillations are observed with a higher S/N ratio compared to
the resistivity and magnetization, which is due to the absence of background in the elastic property measurements.
Since the interaction between ultrasound and carriers occurs only in the vicinity of the Fermi surface, the attenuation peaks correspond exact level crossing points between Fermi level and Landau levels \cite{LK_form}. Hence, ultrasonic measurements can be a powerful probe for the investigation of quantum oscillations. Comparing Figs. 8(a), (b), and (c), we can see that the peaks of $\rho_{xx}$ and $dM/dB$ correspond the depopulations of the Landau level.
We recognize clear anomalies at approximately 18 and 30 T in $\Delta \alpha$ and $\Delta v/v_0$, 
and no noticeable structure was observed in higher fields,
which is consistent with the resistivity and magnetization measurements.

\begin{figure}
\centering
\includegraphics[width=7.5cm]{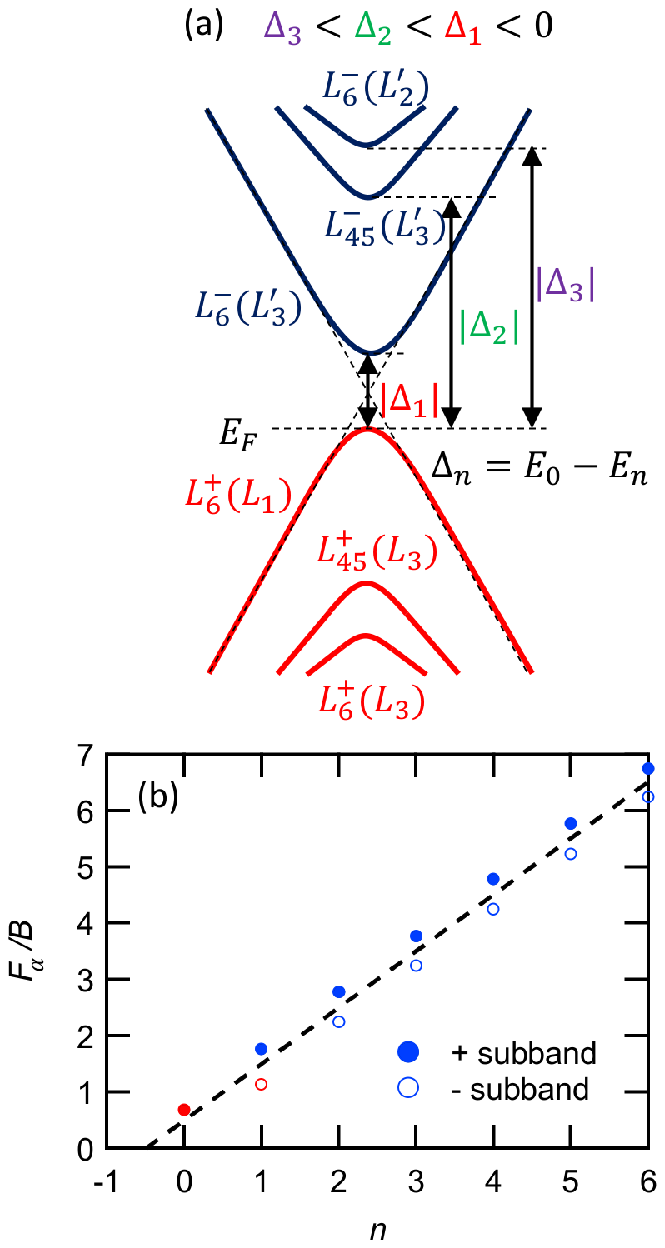}
\caption{
(a) Schematic energy band structure of PbTe at the $L$ point.
The carrier type is assumed to be a hole, and $\Delta_{1\textendash3}$ is the energy difference
from the top of the valence band ($E_0$), defined as $\Delta_n=E_0-E_n$.
(b) The Landau-level fan diagram obtained from the SdH oscillations in the magnetoresistance.
Closed and open symbols represent $+$ and $-$ subbands, respectively.
Blue and red symbols were obtained from measurements in static fields up to 14 T
and pulsed magnetic fields up to 55 T, respectively.
The broken line represents the averaged $F_\alpha/B$ of $+$ and $-$ subbands.
\label{fig10}}
\end{figure}

From these results,
we identified the two quantum oscillations at approximately 18--20 T and 30--32 T,
which indicate the depopulations of the Landau subbands from the Fermi level at these magnetic fields.
These values agree well with the expected values shown in Fig. \ref{fig7}.
Therefore, we can restrict the possible $M_{ZC}$ to 0.52 or 1.48.
In both cases, the system enters the spin-polarized quantum limit state
in which all conduction carriers are confined into the lowest $0-$ subband.
The difference between these two cases is the order of the spin $+$ and $-$.
To distinguish these cases experimentally,
we should perform a measurement to identify the spin corresponding to each subband.
Because such a measurement is not available at the current stage,
we finally adopt a recent theoretical suggestion.
According to the multi-band $\bm{k}\cdot \bm{p}$ theory,
the analytic forms of $M^{\parallel}_{ZC}$ and $M^{\perp}_{ZC}$
are represented by \cite{Hayasaka}:
\begin{align}
M^{\parallel}_{ZC}&=\dfrac{1-\lambda|X|^2+\lambda'|Y|^2}{1+\lambda|X|^2+\lambda'|Y|^2},\label{eq_ZC_parallel}\\
M^{\perp}_{ZC}&=\dfrac{|1-\lambda'YZ^*|}{\sqrt{(1+\lambda'|Z|^2)(1+\lambda|X|^2+\lambda'|Y|^2)}}.\label{eq_ZC_perp}
\end{align}
In this formulation, the carrier type is assumed to be a hole,
and the suffixes $\parallel$ and $\perp$ indicate the magnetic field directions parallel
and perpendicular to the $\langle111\rangle$ direction, respectively.
$\lambda=\Delta_1/\Delta_2$ and $\lambda'=\Delta_1/\Delta_3$,
and $\Delta_n=E_0-E_n$ ($n=1,2,3$) is defined as the energy difference
from the top of the $L_6^+(L_1)$ valence band as shown in Fig. \ref{fig10}(a).
$X$, $Y$, and $Z$ are complex constants determined by the degree of interband couplings via spin-orbit interaction.
Note that the contributions from the lower bands [$L_{45}^+(L_3)$ and $L_{6}^+(L_3)$]
having the same parity as the valence band vanish,
and we can consider only upper bands [$L_{6}^-(L'_3)$, $L_{45}^-(L'_3)$, and $L_{6}^-(L'_2)$] with different parities.
Because $\Delta_{1\textendash3}<0$ in the case of PbTe, $\lambda, \lambda'>0$ is satisfied.
Thus, the denominators are \textit{always} larger than the numerators in Eqs. (\ref{eq_ZC_parallel}) and (\ref{eq_ZC_perp}),
from which it follows that $M_{ZC}<1$ for PbTe.
Thus, we conclude that $M_{ZC}=0.52$ is realized in our \#V sample.
This value lies between that of InSb ($M_{ZC}=0.36$) and bismuth ($M_{ZC}=0.9$--$1.0$) \cite{Fuseya_Dirac}.
On the other hand, SnTe and pressurized PbTe, in which $L_{6}^-(L'_3)$ and $L_{6}^+(L_1)$
are exchanged by band inversion, $M_{ZC}>1$ is shown by similar consideration
using Eqs. (\ref{eq_ZC_parallel}) and (\ref{eq_ZC_perp}) \cite{Hayasaka}.
Thus, we can expect $M_{ZC}=1$, namely, ideal 3D Dirac electron system at the band inversion point.
The magnitude relation between $M_{ZC}$ and 1 is determined only by the order of the energy bands
at the $L$ point.
Introduction of interaction may opens a gap even if $M_{ZC}=1$.
This gap, however, will not affect the argument of $M_{ZC}$ as seen in the case of elemental bismuth \cite{Fuseya_Dirac}.
We note that obtained ZC ratio $M_{ZC}=0.52$ is considerably smaller than theoretical expectation ($M_{ZC}=0.834$) \cite{Hayasaka}. Based on Eqs. (\ref{eq_ZC_parallel}) and (\ref{eq_ZC_perp}), one possibility of this mismatch is the deviation of the degree of inter-band couplings ($X$, $Y$, and $Z$) and energy difference ($\Delta_{1-3}$) assumed in the theoretical calculation from real values. The reason of this discrepancy should be clarified in future study.
Figure \ref{fig10}(b) shows the Landau-level fan diagram constructed from
the $\rho_{xx}$ measurements up to 55 T.
The blue symbols denote measurements in static field up to 14 T,
and the red symbols denote measurements in pulsed magnetic fields.
The dashed line are the averaged $F_\alpha/B$ of the $+$ and $-$ subbands,
which has an $x$-intercept of $\sim$$-0.47$.

\begin{figure}
\centering
\includegraphics[width=7.5cm]{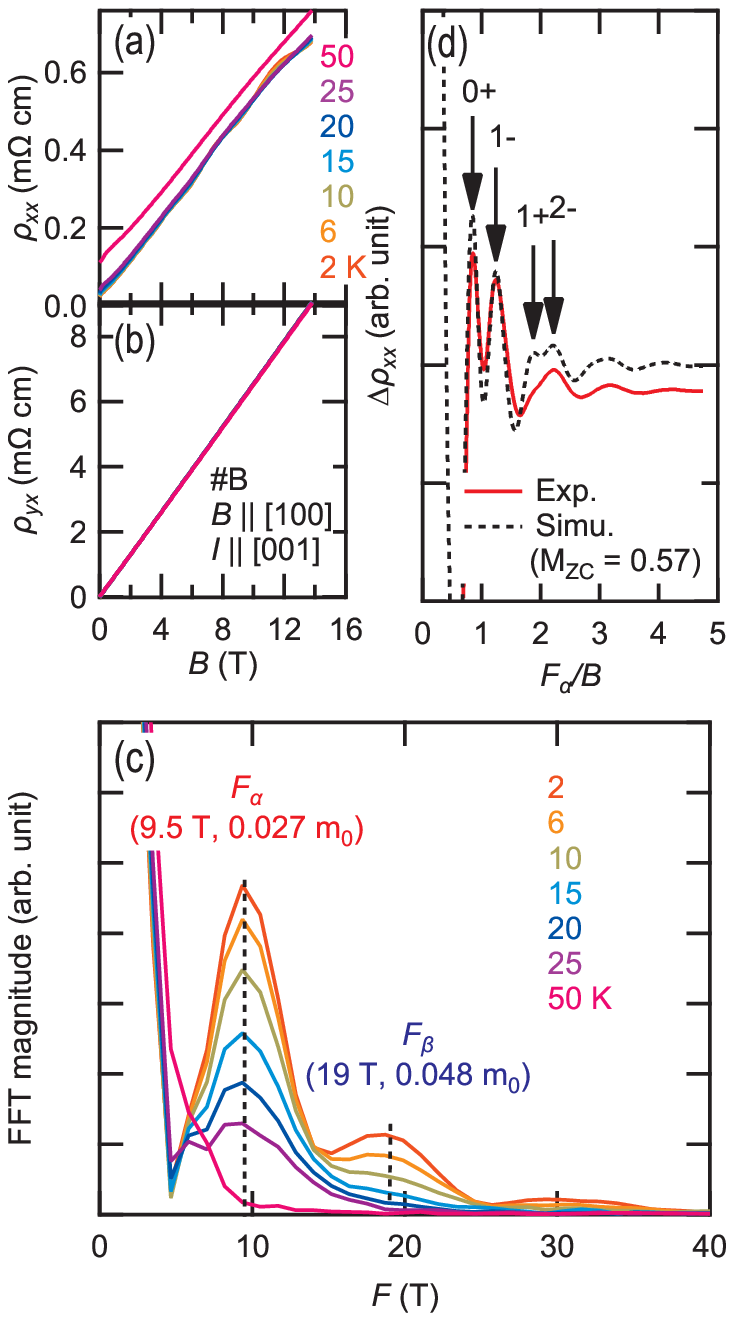}
\caption{
Magnetic field dependence of (a) $\rho_{xx}$ and (b) $\rho_{yx}$ up to 14 T in the \#B sample.
(c) FFT spectra at several temperatures.
(d) Comparison between experimental oscillation $\Delta \rho_{xx}$ (red) and calculation based on Eq. (\ref{eq_LK})
assuming $M_{ZC}=0.57$ (black).
\label{fig9}}
\end{figure}

In the following, we mention the transport properties of another sample made by the Bridgmann method (referred as the
\#B sample), which has a carrier density of $0.99\times10^{18} $cm$^{-3}$ lower than the \#V sample.
The RRR [$\rho$(300 K)/$\rho$(2 K)] of the \#B sample was found to be 384, which is slightly lower than that of the \#V sample.
The metallic behavior over the temperature range from 2 K to 300 K was confirmed to be identical with that of the \#V sample.
Figure \ref{fig9}(a) shows $\rho_{xx}$ up to 14 T at several temperatures.
The magnetoresistance is approximately linear unlike that of the \#V sample shown in Fig. \ref{fig1}(b).
Further, in the \#B sample, SdH oscillation was observed over a wide temperature and magnetic field range, which reflects
the low effective mass and high mobility of carriers in the \#B sample.
Figure \ref{fig9}(b) shows $\rho_{yx}$ up to 14 T.
The slope of $\rho_{yx}$ is steeper than that of the \#V sample shown in Fig. \ref{fig1}(c)
due to the lower carrier density of the \#B sample, while the linearity and sign of $\rho_{yx}$ are identical with those
of the \#V sample.
In, Fig. \ref{fig9}(b), the traces at different temperature almost overlap with each other, which is consistent with the result on the \#V sample shown in Fig. \ref{fig1}(c).

We then focus on the SdH oscillation on $\rho_{xx}$ in detail.
Figure \ref{fig9}(c) shows the FFT spectra at several temperatures from 2 K to 50 K.
We can identify two obvious peaks, at 9.5 and 19 T, and assume that the fundamental
frequency $F_\alpha$ is 9.5 T,
and the other frequency $F_\beta$ with $9.5\times 2=19$ T is
the second harmonic due to the Zeeman-splitting.
In fact, the estimated cyclotron masses from $F_\beta$ (0.048 $m_0$) are approximately two times those from
$F_\alpha$ (0.027 $m_0$).
We performed the peak-position fitting to $\Delta \rho_{xx}$ on the basis of the LK formula and successfully reproduced the experimental
result by taking $M_{ZC}=0.57$.
Also in the present case, we cannot distinguish between $M_{ZC}=0.57$ and 1.43 because the corresponding spins
of the Landau levels cannot be distinguished experimentally.
We assume, however, that $M_{ZC}=0.57$ by a similar assumption to that made in the discussion on the \#V sample.
This value is close to that in the \#V sample, and thus we can assume that
the carrier-density dependence of $M_{ZC}$ is sufficiently weak at least from (0.99--3.7)$\times 10^{18}$
cm$^{-3}$.
These results indicate that the criterion for ``Diracness'' based on the ZC ratio is less sensitive
to the position of the Fermi level from the band-crossing point at least up to this carrier density.

As mentioned in Sec. \ref{sec_rd}, the $R_S$ term in the LK formula changes its sign at $M_{ZC}=0.5$,
which alters the peak and dip structures in quantum oscillations.
Therefore, we have to be careful when evaluating the Berry's phase from the Landau-level fan diagram analysis of the quantum oscillations
that do not show spin-splitting.
Because topological materials of recent interest have a narrow or zero gap at the relevant band,
the condition $M_{ZC}\sim1$ may be realized as suggested through the standard two-band $\bm{k}\cdot \bm{p}$ theory \cite{Hayasaka}.
As demonstrated in this study, experimental determination of $M_{ZC}$ requires data on quantum oscillation showing clear Zeeman-splitting up to the quantum limit state
and their careful analyses.
In this context, PbTe is an ideal material to study the ``Diracness'' via evaluation of $M_{ZC}$.
Further experimental studies on Sn-substituted and/or pressurized PbTe will clarify the role of ``Diracness'' in the magneto-transport properties of this class of materials.

\section{Conclusions}
We investigated the electric transport, magnetization, and elastic properties of PbTe
and observed clear quantum oscillations in resistivity, magnetization, ultrasonic attenuation,
and sound velocity.
We identified a large second harmonic in the FFT spectra of the quantum oscillations, and pointed out that
this is firm evidence for prominent Zeeman-splitting in PbTe.
The simple band structure and distinct Zeeman-splitting in PbTe
enable us to validate the evaluation of the ZC ratio.
By numerical simulation based on the Lifshitz--Kosevich formula,
we consistently explained the oscillation structure and determined the ZC ratio as 0.52
in pristine PbTe.
The Landau level indices are unambiguously determined by measurements in pulsed high magnetic fields up to 55 T,
which takes PbTe to the spin-polarized quantum limit state above $\sim$$30$ T.
From the angular dependence of the Shubnikov--de Haas oscillations, we confirmed that
the ZC ratio is almost independent of the field direction, which is consistent with
the theoretical prediction.
It is important to further study the magneto-transport features on the systems
where the ZC ratio is tuned to unity by chemical substitution or pressure.

\begin{acknowledgments}
We thank Y. Fuseya for valuable discussions and comments.
The angle-resolved resistivity measurements were carried out using facilities of the Electromagnetic Measurements Section, The Institute for Solid State Physics, The University of Tokyo.
This work is supported by PRESTO, JST (No. JPMJPR16R2), Grant-in-Aid for Young Scientists A (No. 16H06015), and the Asahi Glass Foundation.
K. A. was supported by Grant-in-Aid for JSPS Research Fellow (16J04781).
\end{acknowledgments}

\bibliography{ref}

\end{document}